\documentclass[11pt, a4paper]{article}

\usepackage{url}
\usepackage{graphicx}
\usepackage{textcomp}
\usepackage{authblk}
\usepackage{amssymb}
\usepackage{amsthm}
\usepackage{ulem}
\usepackage{tabularx}

\begin{document}

\title{Tunka-Rex: a Radio Antenna Array for the Tunka Experiment}

\author[1]{F.G.~Schr\"oder\textsuperscript{\footnotesize{*}}}
\author[2]{D.~Besson}
\author[3]{N.M.~Budnev}
\author[3]{O.A.~Gress}
\author[1]{A.~Haungs}
\author[1]{R.~Hiller}
\author[3]{Y.~Kazarina}
\author[4]{M.~Kleifges}
\author[5]{A.~Konstantinov}
\author[5]{E.E.~Korosteleva}
\author[1]{D.~Kostunin}
\author[4]{O.~Kr\"omer}
\author[5]{L.A.~Kuzmichev}
\author[3]{R.R.~Mirgazov}
\author[3]{A.~Pankov}
\author[5]{V.V.~Prosin}
\author[6]{G.I.~Rubtsov}
\author[4]{C.~R\"uhle}
\author[3]{V.~Savinov}
\author[2]{J.~Stockham}
\author[2]{M.~Stockham}
\author[3]{E.~Svetnitsky}
\author[7]{R.~Wischnewski}
\author[3]{A.~Zagorodnikov}

\affil[1]{\small{Institut f\"ur Kernphysik, Karlsruhe Institute of Technology (KIT), Germany}}
\affil[2]{Department of Physics and Astronomy, University of Kansas, USA}
\affil[3]{Institute of Applied Physics ISU, Irkutsk, Russland}
\affil[4]{Institut f\"ur Prozessdatenverarbeitung und Elektronik, Karlsruher Institut f\"ur Technologie (KIT), Deutschland}
\affil[5]{Skobeltsyn Institute of Nuclear Physics MSU, Moskau, Russland}
\affil[6]{Institute for Nuclear Research of the Russlandn Academy of Sciences, Moskau, Russland}
\affil[7]{DESY, Zeuthen, Deutschland}

\date{\normalsize{The following article has been accepted by AIP Conference Proceedings.}\\Proceedings of ARENA 2012, Erlangen, Germany\\ \small{to appear on \url{http://proceedings.aip.org/}}}

\maketitle

\begin{abstract}
Tunka-Rex, the Tunka radio extension, is an array of 20 antennas at the Tunka experiment close to Lake Baikal in Siberia. It started operation on 08 October 2012. The antennas are connected directly to the data acquisition of the Tunka main detector, a $1\,$km\textsuperscript{2} large array of 133 non-imaging photomultipliers observing the Cherenkov light of air showers in dark and clear nights. This allows to cross-calibrate the radio signal with the air-Cherenkov signal of the same air showers - in particular with respect to the energy and the atmospheric depth of the shower maximum, $X_\mathrm{max}$. Consequently, we can test whether in rural regions with low radio background the practically achievable radio precision comes close to the precision of the established fluorescence and air-Cherenkov techniques. At a mid-term perspective, due to its higher duty-cycle, Tunka-Rex can enhance the effective observing time of Tunka by an order of magnitude, at least in the interesting energy range above $100\,$PeV. Moreover, Tunka-Rex is very cost-effective, e.g., by using economic Short Aperiodic Loaded Loop Antennas (SALLAs). Thus, the results of Tunka-Rex and the comparison to other sophisticated radio arrays will provide crucial input for future large-scale cosmic-ray observatories, for which measurement precision as well as costs per area have to be optimized. In this paper we shortly describe the Tunka-Rex setup and discuss the technical and scientific goals of Tunka-Rex.
\end{abstract}

\section{Introduction}
More than 100 years after their discovery \cite{Hess1912}, cosmic rays are still a useful tool to answer open questions in astro and particle physics. What are the sources of the ultra-high energy cosmic ray particles and how are they accelerated? What is the physics of particle collision in the energy range beyond the center-of-mass energies reachable by LHC? Can the measurement of extensive air-showers give us new insights in the underlying particle physics? Answering this questions requires cosmic-ray observatories for air showers above $10^{17}\,$eV, which can surpass the already available measurements in accuracy and statistics. This in turn requires technical developments to improve the detection methods for air showers - aiming at measurement precision and cost-effectiveness at the same time. Digital antenna arrays are thus a promising option to complement air-shower arrays, since radio measurements allow a high-duty cycle and a are supposed to provide a high precision at the same time.

The principal feasibility of air-shower radio detection was already shown by historic experiments in the 1950s and 1960s \cite{jelley,Allan1971}. About 10 years ago, the LOPES \cite{FalckeNature2005, Schroeder_LOPES_ARENA2012} and CODALEMA \cite{ArdouinBelletoileCharrier2005} experiments revived the radio technique using digital data acquisition and computer-based analysis procedures. The experience gained in these experiments is now used in state-of-the-art experiments like AERA \cite{Melissas_AERA_ARENA2012}, and Tunka-Rex, the radio extension of the Tunka experiment in Siberia, close to lake Baikal. Tunka-Rex profits also from developments made for AERA, in particular the \emph{Offline} analysis-software framework of the Pierre Auger Observatory \cite{AbreuRadioOffline2010} will also be used for Tunka-Rex.

The main goal of Tunka-Rex is to test whether radio measurements can indeed provide a measurement precision for the energy and the composition of the primary cosmic rays competitive to the established air-fluorescence and air-Cherenkov techniques. Thus, Tunka-Rex is built on the same site as the Tunka-133 non-imaging photomultiplier array, which detects the air-Cherenkov light of air-showers, and covers the energy range from about $10^{16}\,$eV to $10^{18}\,$eV \cite{TunkaRICAP2011}. By a cross-calibration of coincident events, the radio precision can be determined with an accuracy limited by the air-Cherenkov precision: about $15\,\%$ for the energy and $25\,$g/cm\textsuperscript{2} for the atmospheric depth of the shower maximum, $X_\mathrm{max}$, which is sensitive to the mass and the interaction of the primary particle.

\begin{figure}[t]
\centering
\includegraphics[width=0.9\columnwidth]{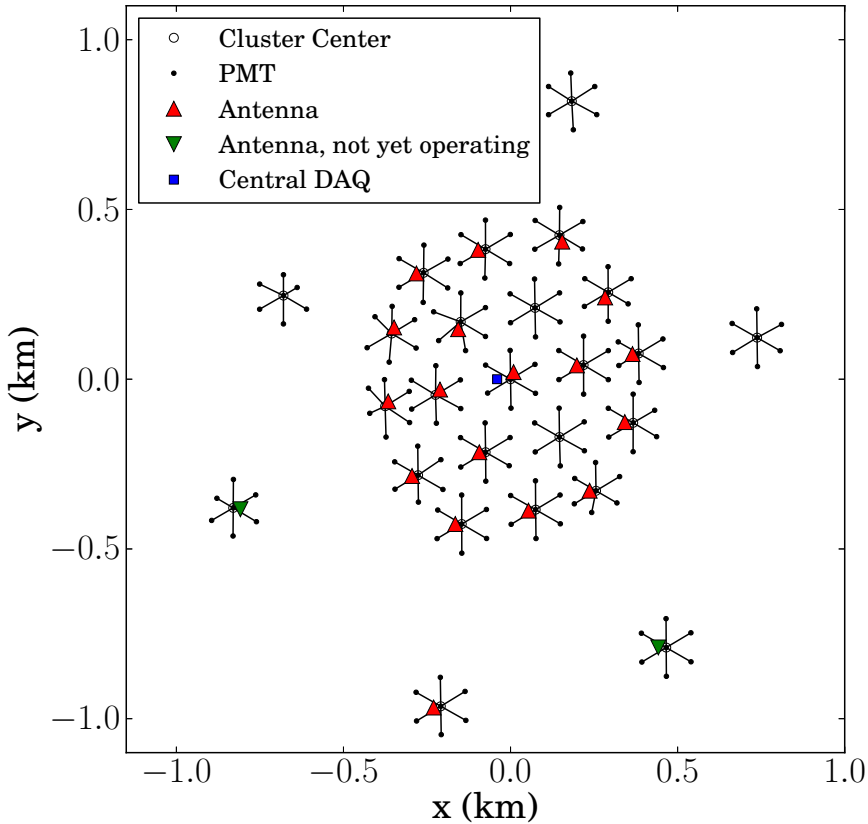}
\caption{Map of the Tunka experiment. The non-imaging photo-multiplier (PMT) array is organized in 25 clusters with 7 PMTs each. Each cluster center features its own digital data acquisition, which is used for the Tunka-Rex antennas, too. The measured data of each cluster are transferred via optical fibers to the central data acquisition (DAQ).} \label{fig_map}
\end{figure}

\section{Tunka-Rex Setup}
The Tunka-Rex array consists of 20 antennas with a typical spacing of approximately $200\,$m, and covers an area of about $1\,$km\textsuperscript{2} (figure \ref{fig_map}). The pattern follows the hexagonal structure of the Tunka PMT array which consists of 19 clusters with 7 PMTs each, and 6 satellite clusters to increase the area and thus the statistics for high energy events above $10^{17}\,$eV. Each cluster features its own local data-acquisition and trigger. The data from the different clusters are transferred via optical fibers to a central data-acquisition which stores the data. Consequently, merging events from different clusters, data analysis and event reconstruction can be done at any later moment offline.

The radio antennas are connected to the same cluster data-acquisition as the PMTs using free ADC channels on the same electronics board. Thus, the radio signal is read simultaneously with the PMT signal of each shower, which is crucial for performing the cross-calibration between the air-Cherenkov and the radio signal. Moreover, the radio and the air-Cherenkov signal is stored in the same data files which facilitates a hybrid analysis. The technical proof-of-principle was made with a prototype antenna installed in 2009, which detected about 70 radio candidate events in the first year of operation \cite{TunkaRICAP2011, SchroederECRS2012}.

\begin{figure}[t]
\centering
\includegraphics[width=0.9\columnwidth]{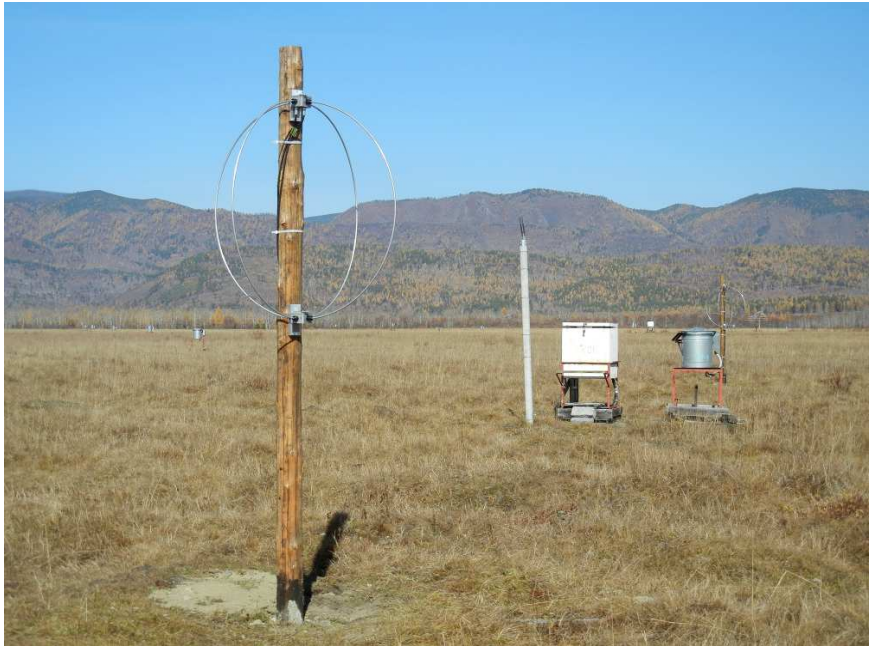}
\caption{Tunka-Rex antenna (SALLA). It is connect to the cluster DAQ (white box) which is close to the one PMT detector.} \label{fig_antenna}
\end{figure}

\subsection{Hardware}
Since the main goal of Tunka-Rex is to demonstrate a high precision for air-shower observables, the analog radio hardware is optimized for measurement accuracy, accepting a slightly higher detection threshold than otherwise possible. As antenna type we have decided for a SALLA \cite{KroemerSALLAIcrc2009} with $120\,$cm diameter. Although the SALLA provides a signal-to-noise ratio worse compared to other antenna types \cite{AbreuAntennaPaper2012}, it has two major advantages: First, it is economic, compact and simple. Second, the gain of the SALLA is relatively independent from environmental conditions. This is important, since the measurement accuracy of pulses with high amplitude is not limited by noise, but by variations of the gain due to environmental changes, e.g., ground conditions \cite{NehlsHakenjosArts2007, SchroederNoise2010}.

Like other modern radio arrays, Tunka-Rex features two orthogonal antennas at each position, to study the polarization of the radio signal. Moreover, a reconstruction of the electric field vector and, thus, the total amplitude is possible when the arrival direction of the radio signal is known \cite{FuchsThesis2012}. Experiments like LOPES, CODALEMA, and AERA decided to align their antennas in east-west, respectively north-south direction. This has the advantage of a slightly lower detection threshold, since the radio signal is for most shower geometries mainly east-west polarized \cite{IsarArena2008, Ardouin2009}, but the disadvantage that in many events the north-south component can not be distinguished from noise. For Tunka-Rex we have decided to rotate the antenna alignment by $45^\circ$, as it is done at LOFAR \cite{NellesARENA_LOFAR2012}. By this we hope to increase the reconstruction accuracy of the electrical field vector, since the signal-to-noise ratio in the two cross-polarized antennas should be roughly equal.

All antennas are connected with RG 213 coaxial cables of about $30\,$m length to the DAQ of the corresponding Tunka cluster. They are placed at a distance of at least $20\,$m to the closest Tunka detector and the DAQ. Thus, possible radio interferences from the PMTs can be distinguished by timing, since the air shower radio pulse is supposed to arrive significantly earlier than any possible disturbance from the PMTs.

In the cluster DAQ, the radio signal is amplified and filtered to a band-width of $30$ to $80\,$MHz. We use steep filters to fulfill the Nyquist sampling theorem, which in our case requires that there any signal or noise at frequencies above $100\,$MHz is strongly suppressed and thus negligible against the signal in the measurement band-width. Therefore, an accurate reconstruction of the original signal with the bandwidth is possible by up-sampling. We have tested with a climate chamber that the gain of the main amplifier is stable within $\pm 1\,$dB, and we plan to test the temperature behavior of the whole system including cables and the low noise amplifier of the antenna. For digitizing we use the Tunka electronics described in reference \cite{TunkaRICAP2011}. The signal is digitized with a 12-bit analog-to-digital converter (type AD9430) using a sampling frequency of $200\,$MHz and a trace length of $5\,$\textmu s.

\begin{table}
\centering
\caption{Tunka-Rex technical data.}
\label{tab_technicalData}
\begin{tabular}{lc} 
\hline
Area& $1\,$km\textsuperscript{2}\\
Number of Antennas& 20\\
Antenna type& SALLA\\
Frequency range& $30-80\,$MHz\\
Trigger and data acquisition& by Tunka-133 PMT array\\
\hline
\end{tabular}
\end{table}

Tunka-Rex will use a reference beacon to monitor and improve the relative timing between different antennas, as LOPES and AERA already do \cite{SchroederTimeCalibration2010, SchroederThesis2011}. For this we will analyze the phasing of continuous sine waves emitted by the beacon. An accurate relative timing between the different antennas is essential -- not only for the reconstruction of the arrival direction. LOPES has shown that the individual antennas can be digitally combined to a radio interferometer, provided that the relative timing is precise and accurate to about $1\,$ns \cite{SchroederTimeCalibration2010}. Moreover, an accurate timing makes it possible to reconstruct the shape of the radio wavefront which is sensitive to the longitudinal shower development and, thus, to the type and mass of the primary particle \cite{Lafebre2010, SchroederIcrc2011}. 

\section{Goals of Tunka-Rex}
The main goal of Tunka-Rex is to test whether the properties of the primary particle can be reconstructed with a precision competitive to other detection technologies. For the arrival direction and the primary energy, LOPES already demonstrated a sufficient precision of $0.65^\circ$, respectively $20\,\%$ \cite{Schroeder_LOPES_ARENA2012}. The best way to reconstruct the composition of the primary particles from radio measurements likely is to statistically analyze the distribution of $X_\mathrm{max}$. LOPES demonstrated that $X_\mathrm{max}$ can be in principle reconstructed from the radio measurements \cite{2012ApelLOPES_MTD, PalmieriARENA2012}, but to our knowledge no radio experiment has yet demonstrated an $X_\mathrm{max}$ precision competitive to air-Cherenkov and air-fluorescence measurements. However, the main reason for the bad $X_\mathrm{max}$ resolution of LOPES seems to be the high ambient noise level there. In rural areas, the ambient noise level usually is much lower \cite{ITU_R_P372}, and at least for the rate of background pulses, this is already confirmed by first measurements at Tunka. Thus, for Tunka-Rex we expect a better $X_\mathrm{max}$ precision, which we will check by a cross-calibration with the radio and air-Cherenkov hybrid events.

Another goal of Tunka-Rex is to improve the understanding of the radio emission by air showers. We will study the properties of the radio signal, in particular the lateral distribution, the shape of the wavefront, the polarization of the radio signal, and the frequency spectrum. Thus, we can cross-check results from other experiments. E.g., historic experiments pointed towards an exponential lateral distribution function (LDF) \cite{Allan1971}, which is also used by LOPES \cite{2010ApelLOPESlateral}, but newer results obtained at AERA indicate that a Gaussian LDF is more appropriate to fit a large distance range \cite{FuchsThesis2012}. The wavefront shape so far has only been studied by LOPES and was found to be approximately conical \cite{SchroederIcrc2011}. The polarization seems to be predominately determined by the geomagnetic effect \cite{KahnLerche1966, IsarArena2008}, but is influenced by the Askaryan effect, too \cite{Askaryan1962, FraenkelECRS2012}. The shape of the frequency spectrum is not yet well known. It is falling towards higher frequencies \cite{NiglFrequencySpectrum2008}, and its slope seems to be sensitive to the longitudinal shower development \cite{GrebeARENA2012}, which provides another method for $X_\mathrm{max}$ reconstruction.

Moreover, we can use Tunka-Rex to test models for the radio emission. A recent comparison of LOPES measurements with REAS 3.11 simulations \cite{LudwigREAS3_2010, LudwigARENA2012} indicates that REAS 3.11 can describe measured radio events -- in contrast to some other radio simulation codes not including the refractive index of air, which changes the coherence conditions for the radio emission \cite{deVries2011}. Due to the lower ambient noise level, we expect that Tunka-Rex can test models to a higher precision than LOPES. In addition, we plan to compare Tunka-Rex measurements to the predictions of several radio simulation codes, e.g., the one described in reference \cite{KalmykovKonstantinov2011}, which will help to further improve the understanding of the physics processed behind the radio signal from air showers.

\section{Conclusion}
Tunka-Rex is based on the success of many other radio arrays for air shower measurements. It makes use of the world-unique opportunity to perform a cross-calibration between air-Cherenkov and radio measurements of the same cosmic ray events. Thus, it will test the expectation whether radio measurements in rural areas really allow a reconstruction precision for the energy and $X_\mathrm{max}$ similar to the precision of the air-Cherenkov measurements. If so, Tunka-Rex can enhance the duty cycle of Tunka for high energy events by an order of magnitude, since it can measure also during days and during almost any weather conditions (except close-by thunderstorms). For this we foresee to trigger Tunka-Rex by the planned scintillator extension of Tunka. Finally, comparing the results of Tunka-Rex with other next generation experiments like AERA will reveal whether a sufficient understanding of the radio signal is already achieved, and under which conditions it makes sense to extend future cosmic ray observatories by digital antenna arrays.

\section*{Acknowledgments}
\small{We acknowledge the support of the Russian Federation Ministry of Education and Science (G/C16.518.11.7051, 14.740.11.0890, 16.518.11.7051, P681, 14.B37.21.0785), the Russian Foundation for Basic research (grants 10-02-00222, 11-02-12138, 12-02-10001, 12-02-91323), the President of the Russian Federation (grant MK-1632.2011.2), the Helmholtz association (grant HRJRG-303).}

\bibliographystyle{unsrt}
\bibliography{arena2012}

\begin{thebibliography}{10}

\bibitem{Hess1912}
V.~F. {Hess}.
\newblock {\"Uber Beobachtungen der durchdringenden Strahlung bei sieben
  Freiballonfahrt}.
\newblock {\em Physikalische Zeitschrift}, 13:1084, 1912.

\bibitem{jelley}
J.~V. {Jelley} et~al.
\newblock {Radio Pulses from Extensive Cosmic-Ray Air Showers}.
\newblock {\em Nature}, 205:327--328, 1965.

\bibitem{Allan1971}
H.~R. {Allan}.
\newblock {Radio Emission From Extensive Air Showers}.
\newblock {\em Progress in Elementary Particle and Cosmic Ray Physics},
  10:171--302, 1971.

\bibitem{FalckeNature2005}
{H.~Falcke et al.~- LOPES Collaboration}.
\newblock {Detection and imaging of atmospheric radio flashes from cosmic ray
  air showers}.
\newblock {\em Nature}, 435:313--316, 2005.

\bibitem{Schroeder_LOPES_ARENA2012}
{F.G.~Schr\"oder et al.~- LOPES Collaboration}.
\newblock {Cosmic Ray Measurements with LOPES: Status and Recent Results}.
\newblock In {\em {Proc. of the ARENA 2012 workshop (Erlangen, Germany)}}, AIP
  Conf. Proc., to be published.

\bibitem{ArdouinBelletoileCharrier2005}
{D.~Ardouin et al.~- CODALEMA Collaboration}.
\newblock {Radio-detection signature of high-energy cosmic rays by the CODALEMA
  experiment}.
\newblock {\em Nucl.~Instr.~and Meth.~A}, 555:148--163, 2005.

\bibitem{Melissas_AERA_ARENA2012}
{M.~Melissas for the Pierre Auger Collaboration}.
\newblock {Recent Developments of the Auger Engineering Radio Array}.
\newblock In {\em {Proc. of the ARENA 2012 workshop (Erlangen, Germany)}}, AIP
  Conf. Proc., to be published.

\bibitem{AbreuRadioOffline2010}
{P.~Abreu et al.~- Pierre Auger Observatory}.
\newblock {Advanced functionality for radio analysis in the Offline software
  framework of the Pierre Auger Observatory}.
\newblock {\em Nucl.~Instr.~and Meth.~A}, 635:92--102, 2011.

\bibitem{TunkaRICAP2011}
{S.F.~Berezhnev et al.~- Tunka Collaboration}.
\newblock {The Tunka-133 EAS Cherenkov light array: Status of 2011}.
\newblock {\em Nucl.~Instr.~and Meth.~A}, 692:98--105, 2012.

\bibitem{SchroederECRS2012}
{F.G.~Schr\"oder et al.~- Tunka-Rex}.
\newblock {Tunka-Rex: a Radio Extension of the Tunka Experiment}.
\newblock {\em {Proc. of 23rd ECRS, Moscow, Russia, to appear in JPCS}}, 2012.

\bibitem{KroemerSALLAIcrc2009}
{O.~Kr\"omer et al.~- LOPES Collaboration}.
\newblock {New Antenna for Radio Detection of UHECR}.
\newblock {\em Proc. of 31st ICRC, {\L}\'{o}d\'{z}, Poland}, page \#1232, 2009.

\bibitem{AbreuAntennaPaper2012}
{P.~Abreu et al.~- Pierre Auger Observatory}.
\newblock {Antennas for the Detection of Radio Emission Pulses from Cosmic-Ray
  induced Air Showers at the Pierre Auger Obervatory}.
\newblock {\em Journal of Instr.}, 7:P10011, 2012.

\bibitem{NehlsHakenjosArts2007}
S.~{Nehls}, A.~{Hakenjos}, M.~J. {Arts}, et~al.
\newblock {Amplitude calibration of a digital radio antenna array for measuring
  cosmic ray air showers}.
\newblock {\em {Nucl.~Instr.~and Meth.~A}}, 589(3):350 -- 361, 2008.

\bibitem{SchroederNoise2010}
{F.G.~Schr\"oder et al.~- LOPES Collaboration}.
\newblock {On noise treatment in radio measurements of cosmic ray air showers}.
\newblock {\em Nucl.~Instr.~and Meth.~A (ARENA 2010)}, 662, Suppl.
  1:S238--S241, 2012.

\bibitem{FuchsThesis2012}
B.~{Fuchs}.
\newblock {PhD Thesis}, Karlsruhe Institute of Technology (KIT), Germany, 2012.

\bibitem{IsarArena2008}
{P.G.~Isar et al.~- LOPES Collaboration}.
\newblock {\em {Nucl.~Instr.~and Meth.~A; Proc. of ARENA 2008}}, 604:81--84,
  2009.

\bibitem{Ardouin2009}
{D.~Ardouin et al.~- CODALEMA Collaboration}.
\newblock {Geomagnetic origin of the radio emission from cosmic ray induced air
  showers observed by CODALEMA}.
\newblock {\em Astroparticle Physics}, 31(3):192 -- 200, 2009.

\bibitem{NellesARENA_LOFAR2012}
{A.~Nelles et al.~- LOFAR}.
\newblock {Detecting Radio Emission from Air Showers with LOFAR}.
\newblock In {\em {Proc. of the ARENA 2012 workshop (Erlangen, Germany)}}, AIP
  Conf. Proc., to be published.

\bibitem{SchroederTimeCalibration2010}
{F.G.} {Schr\"oder}, T.~{Asch}, L.~{B\"ahren}, et~al.
\newblock {\em Nucl.~Instr.~and Meth.~A}, 615(3):277 -- 284, 2010.

\bibitem{SchroederThesis2011}
{F.G.~Schr\"oder}.
\newblock {PhD Thesis}, Karlsruhe Institute of Technology (KIT), Germany, 2011.
\newblock {digbib.ubka.uni-karlsruhe.de/volltexte/1000022313}.

\bibitem{Lafebre2010}
S.~{Lafebre}, H.~{Falcke}, J.~{H\"orandel}, et~al.
\newblock {Prospects for determining air shower characteristics through
  geosynchrotron emission arrival times}.
\newblock {\em Astroparticle Physics}, 34(1):12--17, 2010.

\bibitem{SchroederIcrc2011}
{F.G.~Schr\"oder et al.~- LOPES Collaboration}.
\newblock {Investigation of the Radio Wavefront of Air Showers with LOPES and
  REAS3}.
\newblock {\em {Proc. of 32nd ICRC, Beijing, China}}, 3:\#0313, 2011.
\newblock {www.ihep.ac.cn/english/conference/icrc2011/paper}.

\bibitem{2012ApelLOPES_MTD}
{W.D.~Apel et al.~- LOPES Collaboration}.
\newblock {Experimental evidence for the sensitivity of the air-shower radio
  signal to the longitudinal shower development}.
\newblock {\em Phys. Rev. D}, 85:071101(R), 2012.

\bibitem{PalmieriARENA2012}
{N.~Palmieri et al.~- LOPES Collaboration}.
\newblock {Reconstructing energy and $X_\mathrm{max}$ of cosmic ray air showers
  using the radio lateral distribution measured with LOPES}.
\newblock In {\em {Proc. of the ARENA 2012 workshop (Erlangen, Germany)}}, AIP
  Conf. Proc., to be published.

\bibitem{ITU_R_P372}
{International Telecommunication Union}.
\newblock {Radio noise}.
\newblock (P.372-10), 2009.
\newblock Recommendation ITU-R P.372-10.

\bibitem{2010ApelLOPESlateral}
{W.D.~Apel et al.~- LOPES Collaboration}.
\newblock {Lateral distribution of the radio signal in extensive air showers
  measured with LOPES}.
\newblock {\em Astroparticle Physics}, 32:294--303, 2010.

\bibitem{KahnLerche1966}
F.~D. {Kahn} and I.~{Lerche}.
\newblock {Radiation from cosmic ray air showers}.
\newblock {\em Proc. of Royal Society of London. Series A, Mathematical and
  Physical Sciences}, 289:206, 1966.

\bibitem{Askaryan1962}
G.~A. {Askaryan}.
\newblock {Excess negative charge of an electron-photon shower and its coherent
  radio emission}.
\newblock {\em Soviet Physics JETP}, 14:441, 1962.

\bibitem{FraenkelECRS2012}
{D.~Fraenkel for the Pierre Auger Collaboration}.
\newblock {Measurements and polarization analysis of radio pulses from
  cosmic-ray-induced air showers at the Pierre Auger Observatory}.
\newblock {\em {Proc. of 23rd ECRS, Moscow, Russia, to appear in JPCS}}, 2012.

\bibitem{NiglFrequencySpectrum2008}
{A.~Nigl et al.~- LOPES Collaboration}.
\newblock {Frequency spectra of cosmic ray air shower radio emission measured
  with LOPES}.
\newblock {\em Astronomy \& Astrophysics}, 488:807--817, 2008.

\bibitem{GrebeARENA2012}
{S.~Grebe for the Pierre Auger Observatory}.
\newblock {Spectral index analysis of the data from the Auger Engineering Radio
  Array}.
\newblock In {\em {Proc. of the ARENA 2012 workshop (Erlangen, Germany)}}, AIP
  Conf. Proc., to be published.

\bibitem{LudwigREAS3_2010}
M.~{Ludwig} and T.~{Huege}.
\newblock {\em Astroparticle Physics}, 34:438--446, 2011.

\bibitem{LudwigARENA2012}
{M.~Ludwig et al.~- LOPES Collaboration}.
\newblock {Comparison of LOPES measurements with CoREAS and REAS 3.11
  simulations}.
\newblock In {\em {Proc. of the ARENA 2012 workshop (Erlangen, Germany)}}, AIP
  Conf. Proc., to be published.

\bibitem{deVries2011}
K.~D. {de Vries}, A.~M. {van den Berg}, O.~{Scholten}, and K.~{Werner}.
\newblock {Coherent Cherenkov Radiation from Cosmic-Ray-Induced Air Showers}.
\newblock {\em Phys. Rev. Let.}, 107:061101, 2011.

\bibitem{KalmykovKonstantinov2011}
N.~Kalmykov and A.~Konstantinov.
\newblock Macroscopic model of radio emission from extensive air showers.
\newblock {\em Physics of Atomic Nuclei}, 74:1019--1031, 2011.

\end{thebibliography}

\let\thefootnote\relax\footnotetext{\textsuperscript{\footnotesize{*}} Corresponding author\\Email address: frank.schroeder@kit.edu}

\end{document}